\documentclass[final,times,twocolumn]{elsarticle}
\usepackage{graphicx,amssymb,amsmath}
\usepackage{multirow,dcolumn,bm,latexsym,soul}
\newcolumntype{K}[1]{>{\centering\arraybackslash}p{#1}}

\journal{Physics of the Dark Universe}
\begin{document}
\newcommand{\be}{\begin{equation}}
\newcommand{\ee}{\end{equation}}
\newcommand{\bq}{\begin{eqnarray}}
\newcommand{\eq}{\end{eqnarray}}
\newcommand{\bw}{\begin{widetext}}
\newcommand{\ew}{\end{widetext}}
\newcommand{\bsq}{\begin{subequations}}
\newcommand{\esq}{\end{subequations}}
\newcommand{\bc}{\begin{center}}
\newcommand{\ec}{\end{center}}
\begin{frontmatter}

\title{Low redshift constraints on scale-covariant models}
\author[inst1,inst2]{C. J. A. P. Martins\corref{cor1}}\ead{Carlos.Martins@astro.up.pt}
\author[inst1,inst3]{J. S. J. S. Oliveira}\ead{up201804854@fc.up.pt}
\author[inst1,inst3]{D. A. R. Pinheiro}\ead{up201805213@fc.up.pt}
\address[inst1]{Centro de Astrof\'{\i}sica da Universidade do Porto, Rua das Estrelas, 4150-762 Porto, Portugal}
\address[inst2]{Instituto de Astrof\'{\i}sica e Ci\^encias do Espa\c co, CAUP, Rua das Estrelas, 4150-762 Porto, Portugal}
\address[inst3]{Faculdade de Ci\^encias, Universidade do Porto, Rua do Campo Alegre 687, 4169-007 Porto, Portugal}

\cortext[cor1]{Corresponding author}

\begin{abstract}
The search for a physical model which explains the observed recent acceleration of the universe is a compelling task of modern fundamental cosmology. Recently Fernandes \textit{et al.} presented low redshift observational constraints on a scale invariant model by Maeder. Phenomenologically this can be interpreted as a bimetric theory with a time-dependent cosmological constant. It was shown that a matter density $\Omega_m\sim0.3$ is a poor fit to the data, and the best-fit model would require a fluid with a much smaller density and a significantly positive equation of state parameter. This model is a particular case of an earlier and broader class of models by Canuto \textit{et al.}, which we study here. Specifically, we consider it in two distinct scenarios: as a genuine alternative to $\Lambda$CDM (i.e., without any cosmological constant) and as a parametric extension thereof (where both a cosmological constant and the new mechanism can coexist, and the relative contributions of both are determined by the data). We find that the first scenario can in principle fit the low-redshift data (but a good fit would require values of model parameters, such as the matter equation of state, in conflict with other data), while in the second one the deviation from $\Lambda$CDM is constrained to be small.
\end{abstract}
\end{frontmatter}


\section{Introduction}\label{sect1}

The search for the physical mechanism underlying the observed low redshift acceleration of the universe is the most compelling goal of modern fundamental cosmology. A number of theoretical possibilities can be envisaged in principle, whose observational consequences are being explored. Is dark energy a cosmological constant (i.e. vacuum energy)? If the answer is yes, it is ten to some large power times smaller than our Quantum Field Theory based expectations. If the answer is no, then the Einstein Equivalence Principle must be violated. Either way, new physics is out there, waiting to be discovered; we must search for, identify and characterize this new physics. 

Our goal in this work is to present low redshift background level observational constraints on a class of models which has been claimed to be an alternative to the $\Lambda$CDM paradigm. Recently, Fernandes \textit{et al} \cite{Alternatives} carried out such an analysis for the scale invariant model by Maeder \cite{Maeder}. This model is actually a particular case of a broader class of models, introduced earlier by Canuto \textit{et al.} \cite{Canuto1,Canuto2}: a further specific assumption by Maeder leads to a considerable simplification of the original model. Here we therefore extend the analysis of \cite{Alternatives}, by providing observational constraints for the general model.

In a nutshell, these theories may be motivated by the fact that, although it is well known that the effects of scale invariance are expected to disappear upon the presence of matter (strictly speaking, the presence of particles with non-zero rest masses), one may assume that on large scales, viz. cosmological scales, empty space should still be scale invariant. From a phenomenological point of view, this assumption ultimately leads to a bimetric theory, with a function $\lambda$ playing the role of a scale transformation factor relating the ordinary matter frame to another frame which one assumes to still be scale invariant. The first of these can be thought of as the atomic (or physical) frame, while the second can be thought of as a gravitational frame; in this second frame the ordinary Einstein equations would still hold \cite{Canuto2}. We will study these models both as genuine alternatives to $\Lambda$CDM (in which case they are assumed not to contain a cosmological constant) and as parametric extensions of $\Lambda$CDM (in which case a cosmological constant is allowed, and it and the novel physical mechanism are both are determined by the data).

The plan of the rest of the paper is as follows. We start in Sect. \ref{sect2} with a brief description of our methodology and the data being used. In Sect. \ref{sect3} we introduce the relevant properties of these models and provide a concise review of the model by Maeder and its observational constraints; the relevant part of the mathematical description therein is then extended to the general model Canuto \textit{et al.} in Sect. \ref{sect4}. The observational constraints on the general model are then presented in Sect. \ref{sect5}, under the two assumptions mentioned in the previous paragraph. Finally, in Sect. \ref{sect6} we provide a brief outlook.


\section{Data and methods}\label{sect2}

We take this class of models (to be introduced in the following section) at face value and phenomenologically constrain it through a standard likelihood analysis using low redshift background cosmology data. Specifically, we use two recent and independent datasets. The first is the Pantheon dataset \cite{Scolnic,Riess}, including its covariance matrix. The second is a compilation of 38 Hubble parameter measurements \cite{Farooq}.

We note that since we will only be concerned with low redshift data (specifically $z<2.5$) we ignore the radiation density in what follows. This simplifying assumption has no significant impact in our results. Also, we will work in units where the speed of light is set to $c=1$.

The analyses are done both on a grid and using an MCMC analysis. Since we are only dealing with background cosmology, in most cases the number of free parameters is small enough that there would be no computational need for a full MCMC analysis (although this is indeed needed for the full parameter space that we will be considering). Nevertheless, in several cases we have carried out independent analyses using both techniques, as a means of testing and validating our analysis codes. A further explicit validation test of one of our codes for the supernova data has also been discussed in \cite{Alternatives}.

We follow a standard likelihood analysis \cite{Verde}, with the likelihood defined as
\be
{\cal L}(q)\propto\exp{\left(-\frac{1}{2}\chi^2(q)\right)}\,,
\ee
where $q$ symbolically denotes the parameters for each of the models. Since our two datasets are independent, the total chi-square is the sum of the two, $\chi^2=\chi^2_{SN}+\chi^2_{HZ}$. Our main observable in both cases with be the re-scaled Hubble parameter, which we define here for convenience
\be
E(z)=\frac{H(z)}{H_0}\,;
\ee
we will in general write the Friedmann equation in terms of $E(z)$. Confidence levels are then identified, in terms of the corresponding $\Delta\chi^2$, with standard numerical tools.

Specifically, for the supernovas the Pantheon dataset is a 1048 supernova dataset, containing measurements in the range $0<z<2.3$ \cite{Scolnic}, further compressed into 6 correlated measurements of $E^{-1}(z)$ in the redshift range $0.07<z<1.5$ \cite{Riess}. This provides an effectively identical characterization of dark energy as the full supernova sample, thus making it an efficient compression of the raw data. The chi-square can generically be written
\be
\chi^2_{SN}(q)=\sum_{i,j}\left(E_{obs,i}-E_{model,i}(q)\right)C_{ij}^{-1}\left(E_{obs,j}-E_{model,j}(q)\right)\,,
\ee
where $C$ is the covariance matrix of the dataset. Note that this analysis is independent of the Hubble constant, $H_0$.

Regarding the Hubble parameter dataset, this is a heterogeneous set of 38 measurements some of which come from cosmic chronometers and the rest from baryon acoustic oscillations (BAO). Strictly speaking the BAO measurements rely on some underlying assumptions on a fiducial model, but this model dependence is know not be significant, at least for models close to $\Lambda$CDM (in which case this dependence is at the percent level) \cite{BAO1,BAO2}. The model dependence can be more significant in models with large metric gradients, the typical cases being models with late-time inhomogeneities or those where backreaction is important. In our present case, we therefore expect the dependence not to be significant for the cases with a $\Lambda$CDM limit, though it could conceivably be larger for the ones without this limit---in such case we would have a 'theoretical systematic'. The correlation between these BAO measurements is non-zero but small, so the measurements in the dataset can be assumed to be independent (i.e., the covariance matrix is assumed to be trivial).

On the other hand, in order to do the analysis in terms of $E(z)$ and thus combine the two datasets in the above likelihood one needs to marginalize the Hubble constant, which can be done analytically \cite{Basilakos}. This has the advantage of reducing the parameter space, and also eliminating the need of choosing specific priors on the Hubble constant---a possibly tricky choice given the so-called Hubble tension. With these assumptions, one computes three separate quantities
\bq
A(q)&=&\sum_{i}\frac{E_{model,i}^2(q)}{\sigma^2_i}\\
B(q)&=&\sum_{i}\frac{E_{model,i}(q) H_{obs,i}}{\sigma^2_i}\\
C(q)&=&\sum_{i}\frac{H_{obs,i}^2}{\sigma^2_i}
\eq
where the $\sigma_i$ are the uncertainties in observed values of the Hubble parameter. Then chi-square is
\be
\chi^2(q)=C(q)-\frac{B^2(q)}{A(q)}+\ln{A(q)}-2\ln{\left[1+Erf{\left(\frac{B(q)}{\sqrt{2A(q)}}\right)}\right]}
\ee
where $Erf$ is the Gauss error function and $\ln$ is the natural logarithm.

As a benchmark for the models to be constrained in what follows, we briefly discuss how the above datasets constrain the traditional phenomenological parameterization of Chevallier, Polarski and Linder (henceforth CPL) \cite{Chevallier,Linder}. Here the dark energy equation of state parameter is assumed to have the form
\be
w(z)=\frac{p(z)}{\rho(z)}= w_0+w_a\frac{z}{1+z}\,,
\ee
where $w_0$ is its present value while $w_a$ quantifies its possible evolution in time (or, explicitly, redshift). The $\Lambda$CDM model corresponds to $w_0=-1$ and $w_a=0$. For a flat Friedmann-Lema\^{\i}tre-Robertson-Walker model, the Friedmann equation has the form
\be
E^2(z)=\Omega_m(1+z)^3+(1-\Omega_m)(1+z)^{3(1+w_0+w_a)}\exp{\left(-\frac{3w_az}{1+z}\right)}\,,
\ee
where the matter parameter is defined as usual, $\Omega_m \equiv \kappa \rho_0 / 3 H_0^2$, where $\kappa=8\pi G$ and $\rho_0$ is the present-day critical density. One such analysis is presented in \cite{Alternatives}, and we summarize it in the remainder of this section.

If one assumes a constant equation of state parameter ($w_a=0$), the one-sigma constraints on the two model parameters from the combined data sets are
\bq
\Omega_m&=&0.27\pm0.02\\
w_0&=&-0.92\pm0.06\,,
\eq
which are compatible with $\Lambda$CDM. For the full three-parameter CPL case, the three model parameters from the combined data sets are
\bq
\Omega_m&=&0.26^{+0.03}_{-0.05}\\
w_0&=&-0.92^{+0.09}_{-0.08}\\
w_a&=&0.86^{+0.14}_{-0.24};
\eq
the reduced chi-square at the best fit is $\chi^2_\nu\sim0.9$, so the model is slightly overfitting the data (this behaviour is mainly driven by the Hubble parameter data). The first two of these constraints are compatible with the values for the $w_0$CDM analysis (with naturally larger uncertainties), but there is an apparently clear preference for a positive slope $w_a>0$. However, these $w_a$ constraints strongly depend on the choice of priors (there is no such dependence for the constraints on $\Omega_m$ or $w_0$). The above constraints used the uniform prior on the matter density $\Omega_m=[0.05,0.5]$. As an illustration, if instead one uses the narrower uniform prior $\Omega_m=[0.15,0.45]$, the constraints become
\bq
\Omega_m&=&0.26^{+0.03}_{-0.05}\\
w_0&=&-0.92^{+0.07}_{-0.08}\\
w_a&=&0.74^{+0.21}_{-0.48};
\eq
in other words, there is no impact on the matter density and $w_0$, but there is a significant impact on $w_a$. Breaking these degeneracies requires additional data, for example from cosmic microwave background observations \cite{Planck}.

Our main goal here is to set up a benchmark for the constraining power of these data sets, especially for what concerns the matter density $\Omega_m$, against which to compare the constraints on the alternative models to be discussed in what follows.


\section{Scale invariance: the Maeder model}\label{sect3}

The recently proposed scale invariant model \cite{Maeder} is a particular case of a scale-covariant theories by Canuto \textit{et al.} \cite{Canuto1,Canuto2}, which we will discuss subsequently. With the further assumption of a homogeneous and isotropic universe, the Friedmann, Raychaudhuri, and continuity equations in the general model are \cite{Canuto1,Canuto2}
\bq
\left(\frac{\dot a}{a}+\frac{\dot\lambda}{\lambda}\right)^2+\frac{k}{a^2}&=&\frac{1}{3}(\kappa\rho+\Lambda\lambda^2)\\
\frac{\ddot a}{a}+\frac{\ddot \lambda}{\lambda}+\frac{\dot \lambda}{\lambda}\frac{\dot a}{a}-\frac{{\dot\lambda}^2}{\lambda^2}&=&-\frac{\kappa}{6}(\rho+3p-2\Lambda\lambda^2)\\
{\dot\rho}+3(\rho+p)\frac{\dot a}{a}&=&-(\rho+3p)\frac{\dot \lambda}{\lambda}\,,
\eq
where $k$ is the curvature parameter. These trivially reproduce the standard equations if the function $\lambda(t)$ is set to $\lambda=1$. Note that for a homogeneous and isotropic model $\lambda$ depends only on time, as does the scale factor.

The recent work of Maeder further postulates that the Minkowski metric is a solution of these Einstein equations, which leads to the following consistency conditions \cite{Maeder}
\bq
3\frac{{\dot\lambda}^2}{\lambda^2}&=&\Lambda\lambda^2\\
2\frac{\ddot \lambda}{\lambda}-\frac{{\dot\lambda}^2}{\lambda^2}&=&\Lambda\lambda^2\,,
\eq
which can be used to simplify the Friedmann and Raychaudhuri equations to
\bq
\frac{\dot a^2}{a^2}+2\frac{\dot a}{a}\frac{\dot\lambda}{\lambda} +\frac{k}{a^2}&=&\frac{1}{3}\kappa\rho\\
\frac{\ddot a}{a}+\frac{\dot \lambda}{\lambda}\frac{\dot a}{a}&=&-\frac{\kappa}{6}(\rho+3p)\,;
\eq
note that, superficially, there is now no cosmological constant $\Lambda$; we return to this point in the following section. Moreover, the two consistency conditions imply
\be
\lambda=\sqrt{\frac{3}{\Lambda}}\frac{1}{t}\,.
\ee
If one assumes constant equations of state, $p=w\rho$, together with the solution for $\lambda$, the continuity equation yields
 \be
 \rho\propto (1+z)^{3(1+w)}t^{1+3w}\,.
 \ee
For the particular case of a cosmological constant ($w=-1$) this becomes $\rho\propto t^{-2}$. In other words, the Maeder assumption effectively leads to a model with a time-dependent cosmological constant, but no parametric $\Lambda$CDM limit.

With the aforementioned assumptions, the Friedmann equation for the Maeder model can be written
\begin{equation}
E^2(z,x)=\Omega_m(1+z)^{3(1+w)}x^{1+3w}+\Omega_k(1+z)^2+\frac{\Omega_\lambda}{x}E(z,x)\,,
\label{scalef}
\end{equation}
where the matter parameter $\Omega_m$ has the standard definition, as does the curvature parameter $\Omega_k=-k/(a_0H_0)^2$, and we have also defined an effective parameter
\begin{equation}
\Omega_\lambda=\frac{2}{t_0H_0}\,
\end{equation}
and a dimensionless time $x=t/t_0$, with $t_0$ being the current age of the universe. Note that superficially one might think of $\Omega_\lambda$ as akin to a dark energy parameter, but it is simply related to the present day age of the universe---specifically, it is a measurement thereof, in dimensionless units.

As shown in \cite{Alternatives} this Friedmann equation can be re-written in a simpler form, which we reproduce here for the purpose of comparison to that of the general model discussed in the next section,
\bq
E(z,x)&=&\frac{\Omega_\lambda}{2x}\left[1+\sqrt{1+M(z,x)}\right]\\
\frac{\Omega_\lambda^2}{4}M(z,x)&=&\Omega_m[(1+z)x]^{3(1+w)}+\Omega_k[(1+z)x]^2\,,
\eq
with the relation between redshift and (dimensionless) time being given by
\be
\frac{dx}{dz}=- \frac{x}{1+z}\times\frac{1}{1+\sqrt{1+M(z,x)}}\,,
\ee
together with the initial condition $x=1$ at $z=0$.

In \cite{Maeder} the author claims, from a simple qualitative comparison, that with the choice $\Omega_m=0.3$ the model is in good agreement with Hubble parameter data. This claim has been assessed in Fernandes \textit{et al.} \cite{Alternatives} with a more thorough statistical analysis, whose results we also briefly summarize here.
 
For the $w=0$ case one can write $\Omega_\lambda=1-\Omega_m-\Omega_k$. If one further assumes $\Omega_k=0$, the one-sigma posterior constraint in the matter density is
\begin{equation}
\Omega_m=0.26\pm0.02\,,\quad \chi^2_\nu=1.3\,,
\end{equation}
while if the curvature parameter is allowed to vary with the generous uniform prior $\Omega_k=[-0.2,0.2]$ and marginalized, one finds
\begin{equation}
\Omega_m=0.32\pm0.03\,,\quad \chi^2_\nu=1.2\,;
\end{equation}
it is clear from the structure of the Friedmann equation that the matter and curvature parameters are correlated. Although superficially the best fit values for the matter density are close to those obtained in the previous section for the CPL model, the reduced chi-square at the best fit values is rather poor in both cases, at least in comparison to the value of $\chi^2_\nu=0.9$ obtained for the CPL model therein with the same datasets. This shows that the model is not a good fit to the data, at least by comparison to the standard CPL parameterization.

If instead of assuming a standard matter component one allows its (constant) equation of state parameter $w$ to be a further free parameter, the one-sigma constraints for the $\Omega_k=0$ case are
\bq
\Omega_m &=& 0.06\pm0.02\\ 
w &=&0.60^{+0.16}_{-0.15}\,,
\eq
while if $\Omega_k$ is allowed to vary (with the aforementioned prior) and marginalized one instead gets
\bq
\Omega_m &=& 0.06\pm0.03\\ 
w &=&0.59^{+0.17}_{-0.15}\,.
\eq
In both cases the reduced chi-square is now $\chi^2_\nu=0.8$ (that is, lower than that for the CPL case), so the model is now overfitting the data. Clearly there is a strong degeneracy between the matter density and the equation of state parameter, and the best fit values of both parameters are very far from the standard $\Lambda$CDM ones.


\section{Scale covariance: The Canuto \textit{et al.} model}\label{sect4}

Let us now return to the full equation for the Canuto \textit{et al.} model, introduced in the previous section. Our aim is to constrain this model, using the aforementioned data. We will assume a generic power-law behaviour for the function $\lambda(t)$, specifically
\be
\lambda(t)=\left(\frac{t}{t_0}\right)^p=x^p\,;
\ee
this choice of time dependence is mainly motivated by simplicity, but the specific choice of $\lambda(t_0)=1$ also ensures that $\Lambda$CDM is recovered for $p=0$.

Note that in the particular case of the Maeder model there was no explicit cosmological constant $\Lambda$. In the general case it is still there, so we may expect that there will be two classes of solutions. One has the usual $\Lambda$ providing the acceleration, with the $\lambda$ field providing a further contribution (which is likely constrained to be small); in other words, this will be a parametric extension of $\Lambda$CDM. The other has $\Lambda=0$, meaning that the model will not have a $\Lambda$CDM limit, and the question is then whether the field $\lambda$ can provide an alternative to acceleration in that case. Earlier work \cite{Alternatives} has shown that in the Maeder case this is would only be possible with highly non-standard matter.

In this general case the continuity equation gives
\be
\rho\propto (1+z)^{3(1+w)}t^{-p(1+3w)}\,;
\ee
for the vacuum energy ($w=-1$) case we now have $\rho\propto t^{2p}$, so again a negative $p$ corresponds to a decaying cosmological constant, while for ordinary matter ($w=0$) we have $\rho\propto (1+z)^{3}t^{-p}$. The Friedmann equation is now
\bq
\left(E(z,x)+\frac{p}{2x}\Omega_\lambda\right)^2&=&\Omega_m(1+z)^{3(1+w)}x^{-p(1+3w)}\nonumber \\
& &+\Omega_k(1+z)^2+\Omega_{\Lambda}x^{2p}\,.
\label{scalef2}
\eq
Note that there is a consistency condition
\be
\left(1+\frac{p}{2}\Omega_\lambda\right)^2=\Omega_m+\Omega_k+\Omega_{\Lambda}\,.
\ee
Interestingly, if one chooses $p=-1$ the Maeder model is recovered from these equations with the further assumption that
\begin{equation}
\Omega_\Lambda=\frac{1}{4}\Omega^2_\lambda=\frac{1}{(t_0H_0)^2}\,.
\end{equation}
So it is not strictly true that the cosmological constant vanishes in the Maeder model: although it is not explicit in the Friedmann equation, it is effectively determined by the age of the universe (which would further raise issues of fine-tuning). In passing we also point out that, as previously noted in \cite{Canuto2}, the choice $p=-1$ would be commensurate with Dirac's Large-number Hypothesis \cite{Dirac1,Dirac2}.

Again these general equations can be re-written in the simpler form
\bq
E(z,x)&=&\frac{\Omega_\lambda}{2x}\left[-p+\sqrt{N(z,x)}\right]\\
\frac{\Omega_\lambda^2}{4}N(z,x)&=&\Omega_m(1+z)^{3(1+w)}x^{2-p(1+3w)} \nonumber \\
& & +\Omega_k(1+z)^2x^2+\Omega_\Lambda x^{2(1+p)}\,,
\eq
and the relation between redshift and (dimensionless) time is now given by
\be
\frac{dx}{dz}=- \frac{x}{1+z}\times\frac{1}{\sqrt{N(z,x)}-p}\,,
\ee
with the initial condition still being $x=1$ at $z=0$. One can easily check that the Maeder model equations are recovered in the appropriate limit. For future reference, it must be kept in mind that $\Omega_\lambda$ is a dimensionless measure of the current age of the universe, and it must therefore be a positive quantity.

\begin{figure*}
\begin{center}
\includegraphics[width=\columnwidth]{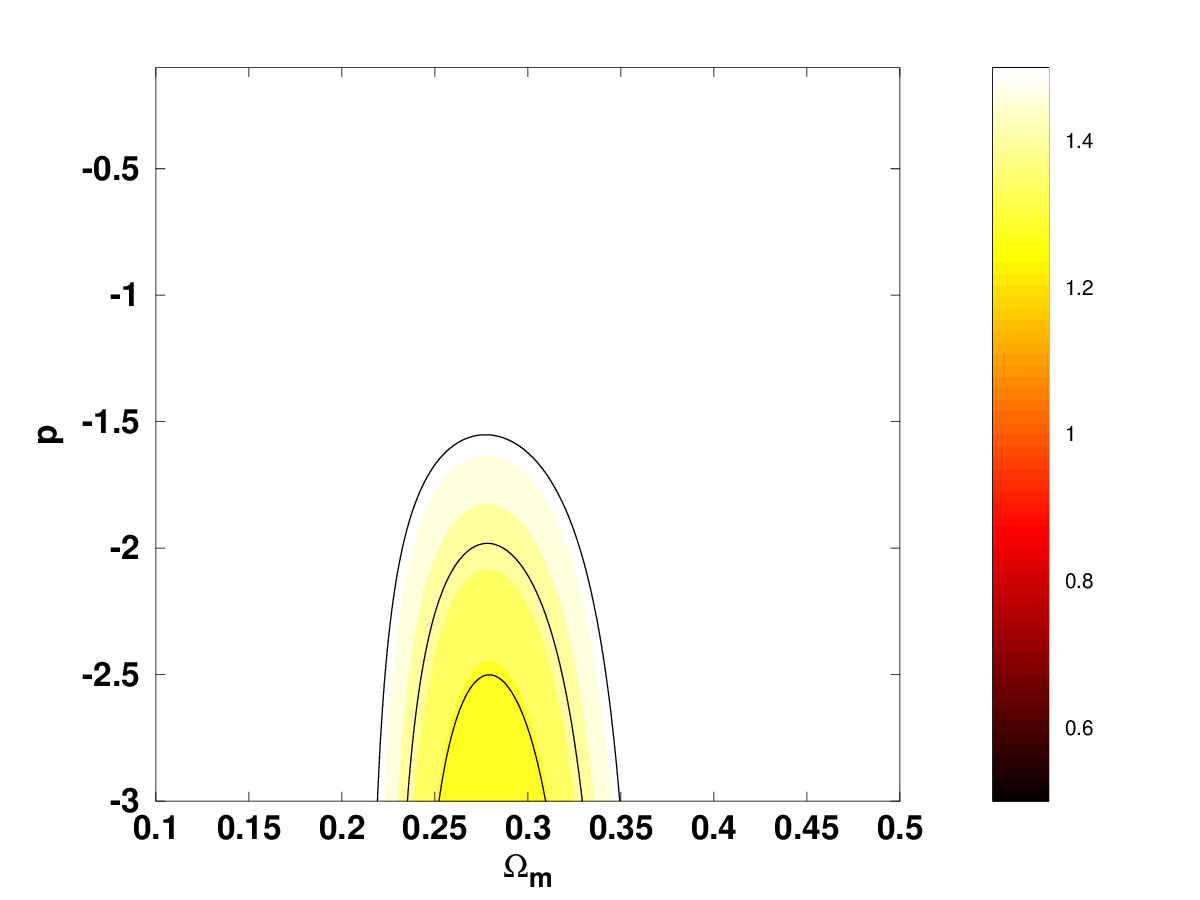}
\includegraphics[width=\columnwidth]{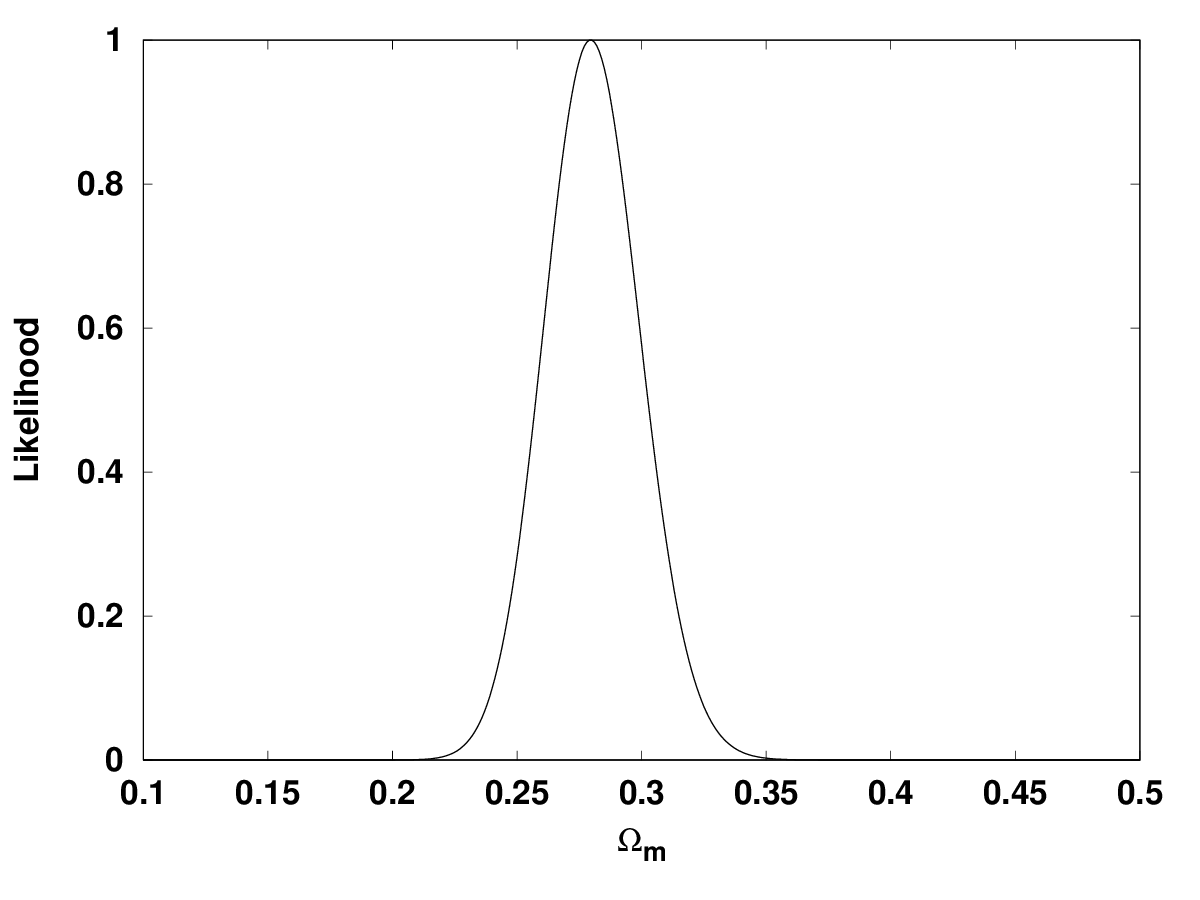}
\includegraphics[width=\columnwidth]{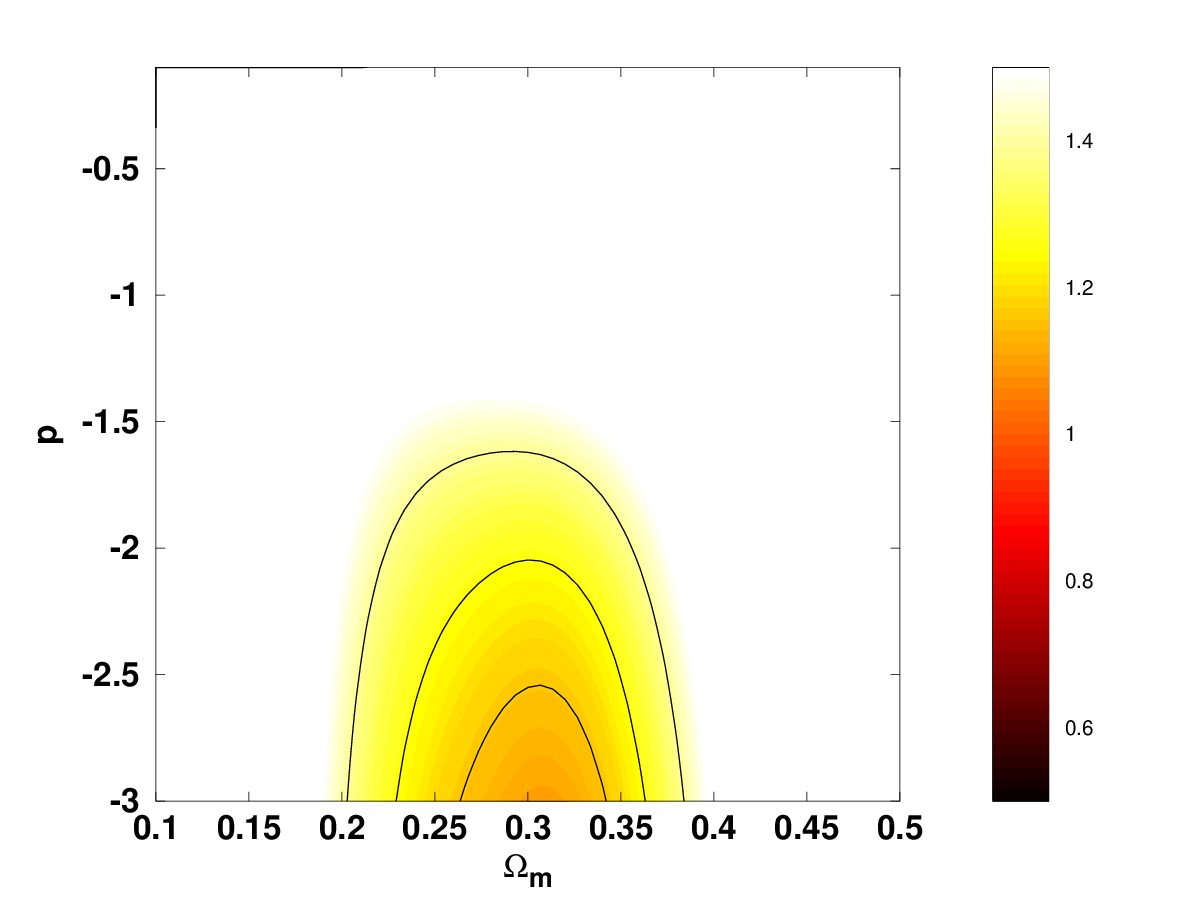}
\includegraphics[width=\columnwidth]{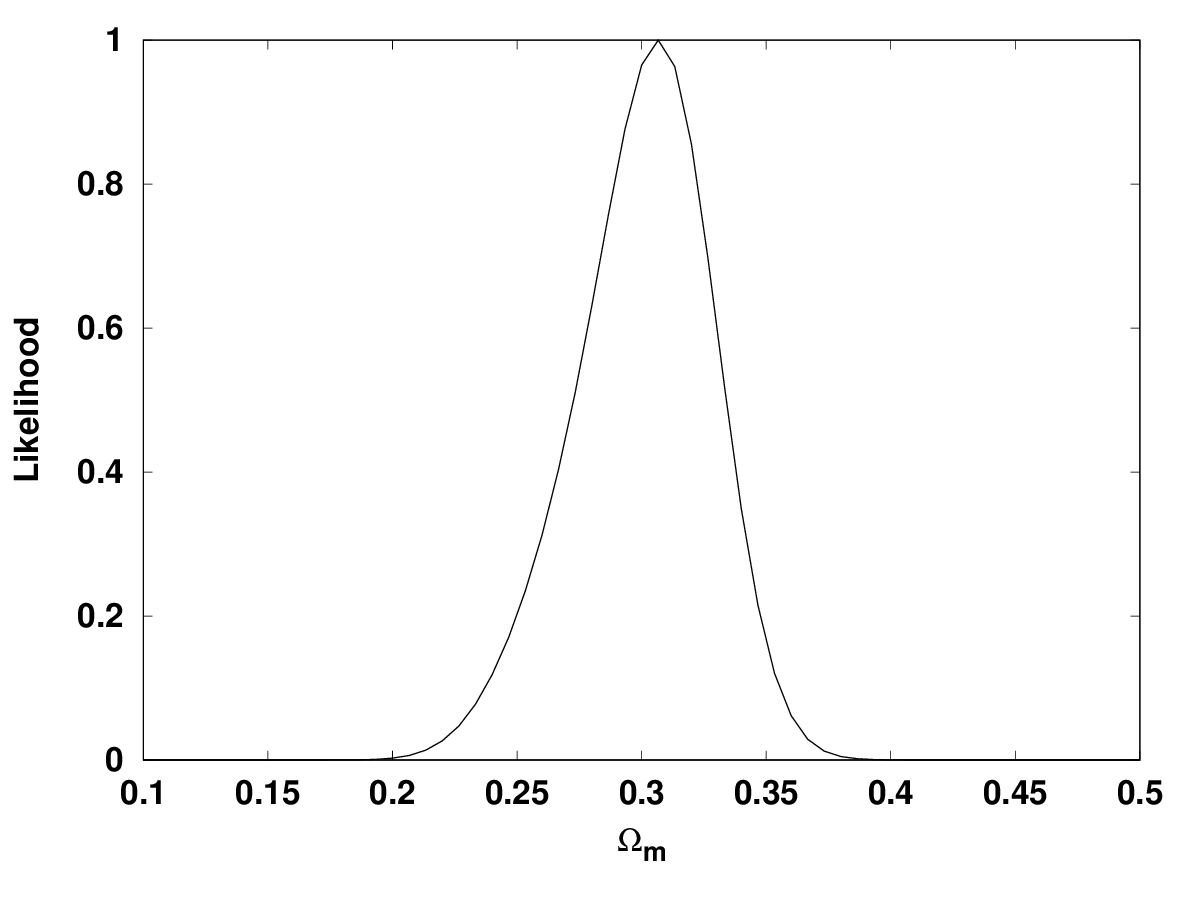}
\end{center}
\caption{\label{fig1}Constraints on the Canuto \textit{et al.} model, with $w=0$ and $\Omega_\Lambda=0$, for universes with $\Omega_k=0$ (top panels) or with $\Omega_k$ marginalized (bottom panels). Left panels: Two-dimensional constraints. The black lines represent the one, two and three sigma confidence levels, and the color map depicts the reduced chi-square at each point in the parameter space, with points with $\chi^2_\nu>1.5$ shown in white. Right panels: One-dimensional posterior likelihood for the matter density.}
\end{figure*}

\section{Constraints on the Canuto \textit{et al.} model}\label{sect5}

As has been previously mentioned, several different scenarios can now be considered, depending on whether or not the model includes a cosmological constant and on the number of free parameters allowed. We separately discuss the two scenarios in each of the following subsections.

\subsection{Without a cosmological constant}

Setting $\Omega_\Lambda=0$ naturally implies that the model will not have a $\Lambda$CDM limit. It follows that the question to be considered is whether the field $\lambda$ can provide an alternative mechanism which accounts for the observed acceleration. In this case the consistency condition, provided by the Friedmann equation and relating the remaining free parameters, is
\begin{equation}
-p\Omega_\lambda=2(1-\sqrt{\Omega_m+\Omega_k})\,,    
\end{equation}
which we will use in this subsection to eliminate $\Omega_\lambda$.

The simplest model in this class stems from assuming a flat universe ($\Omega_k=0$) and ordinary matter with an $w=0$ equation of state parameter\footnote{An early analysis of this simplest case has also been presented by the authors in a recent conference proceedings \cite{Proceedings}.}. In this case we have two free parameters ($\Omega_m,p$) and the case $p=0$ corresponds to an Einstein-de Sitter $\Omega_m=1$ universe, We assume uniform priors in the range $\Omega_m\in[0.1, 0.5]$ and $p\in[-3, 0[$. In this case we again find that this model does not provide a good fit to the data, as shown in the top panels of Fig. \ref{fig1}: while a matter density of
\be
\Omega_m=0.28\pm0.02
\ee
is preferred (which is again similar to the best fit value for the CPL model, given the datasets that we are using), the reduced chi-square of the 3D best fit is quite poor, being always larger than $\chi^2_\nu=1.25$. The lowest value of $p$ allowed by the choice of prior is always the preferred one. Enlarging this prior slightly improves the reduced chi-square but not sufficiently to lead to a good fit: even if we allow the extreme range $p\in[-100, 0[$, the reduced chi-square of the 3D best-fit model is still $\chi^2_\nu=1.06$. In any case, the choice of prior for $p$ has no significant effect on the posterior likelihood for $\Omega_m$.

Allowing for a non-zero curvature has very little effect. While there is a degeneracy between the matter and curvature parameters, even with the rather generous uniform prior $\Omega_k\in[-0.1, 0.1]$ (and keeping the aforementioned priors on the other two parameters) the reduced chi-square of the 3D best fit model only improves to $\chi^2_\nu=1.23$, with one-sigma posterior constraint on the matter density becoming
\be
\Omega_m=0.31\pm0.02\,.
\ee
For comparison, the corresponding constraints are also shown in the bottom panels of Fig. \ref{fig1}.

\begin{figure*}
\begin{center}
\includegraphics[width=\columnwidth]{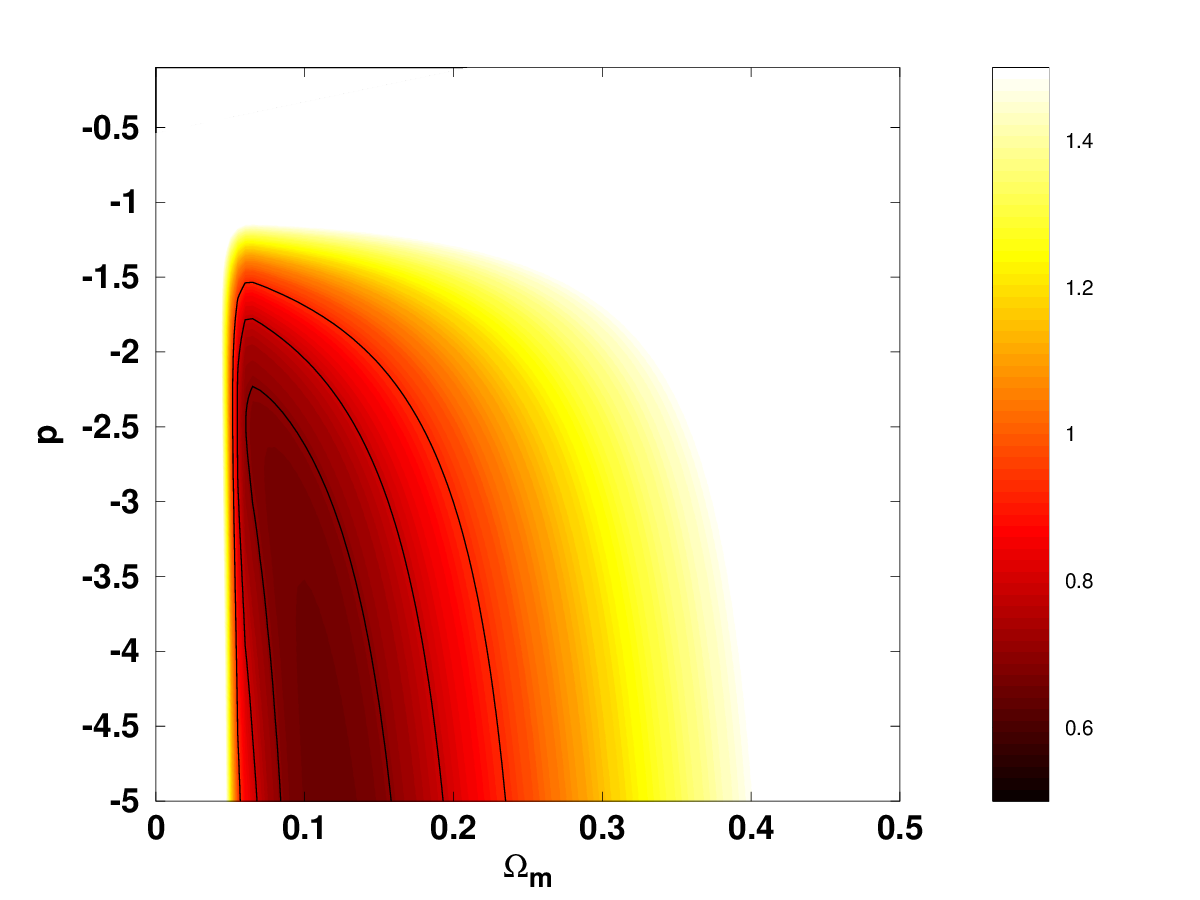}
\includegraphics[width=\columnwidth]{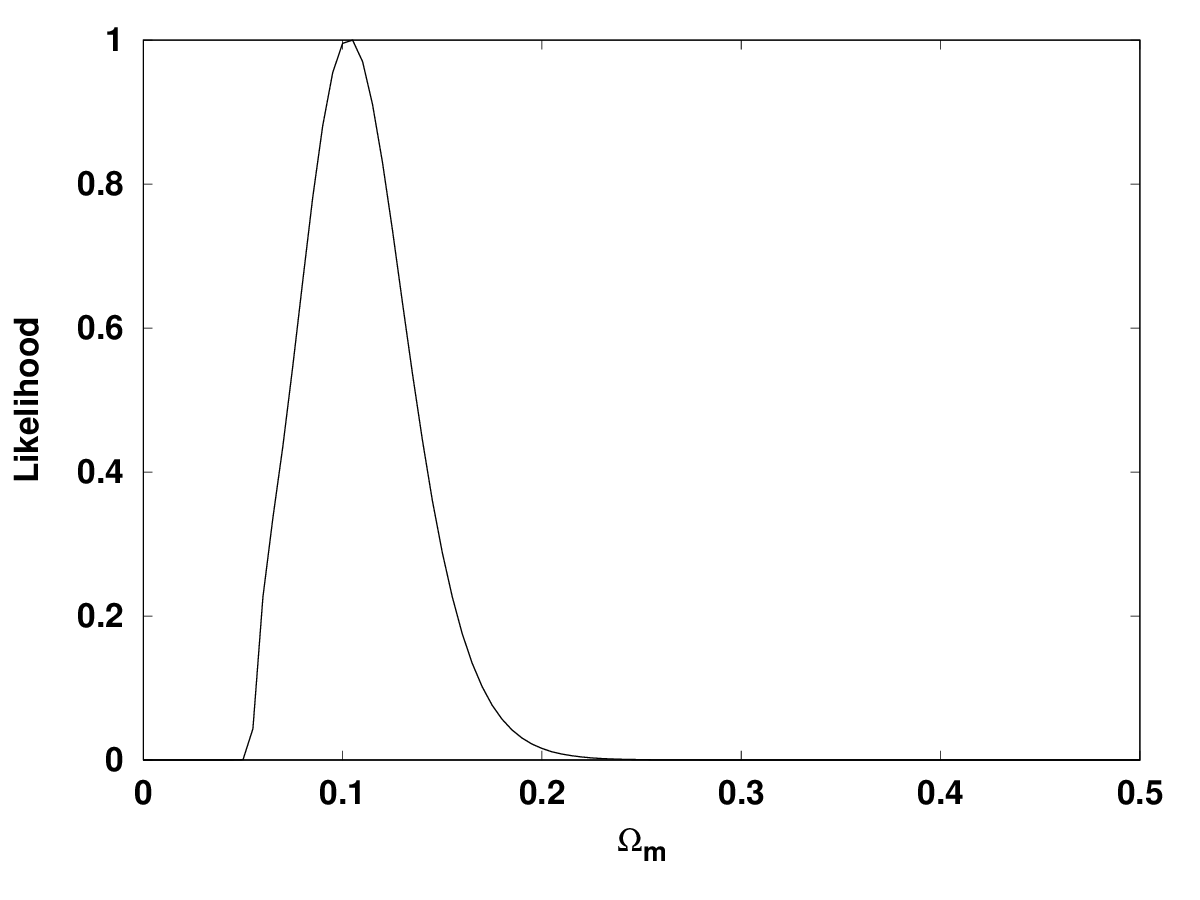}
\includegraphics[width=\columnwidth]{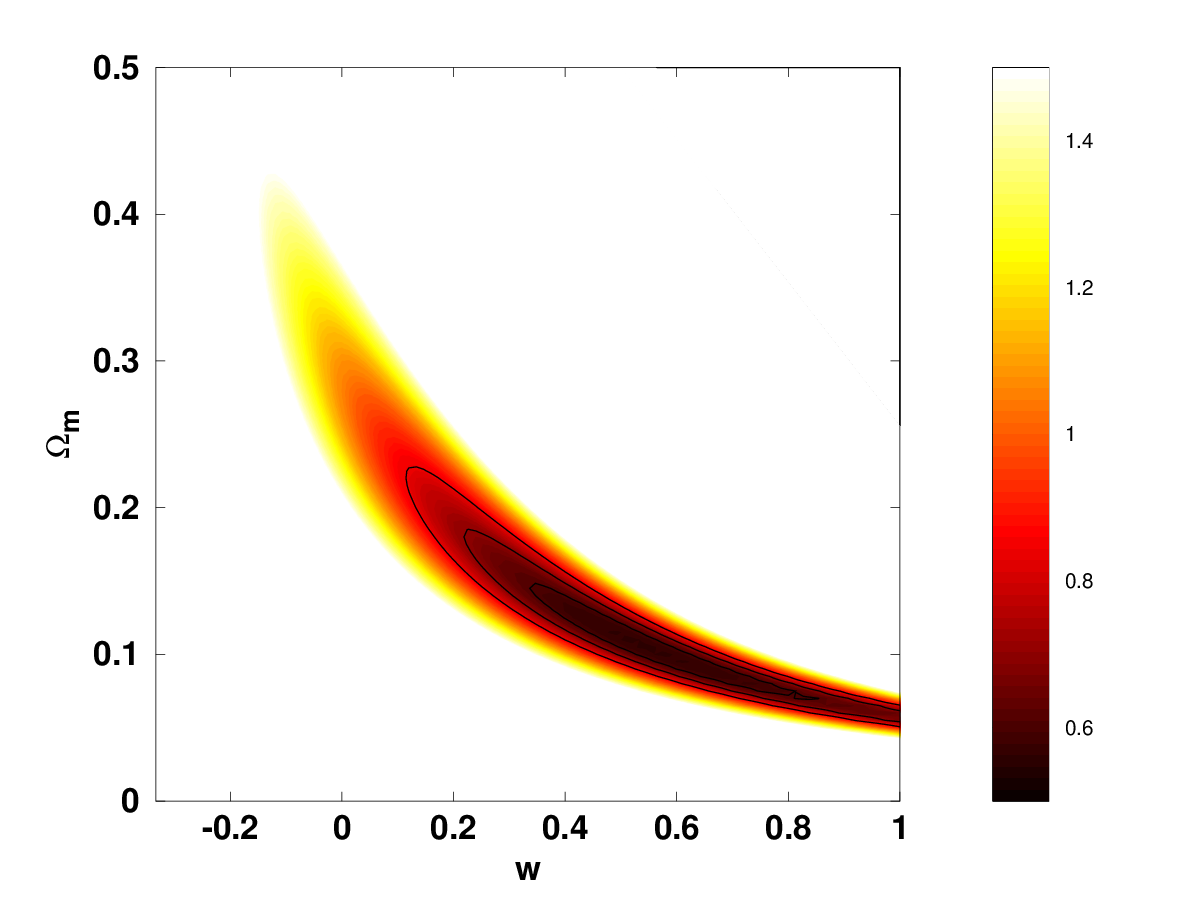}
\includegraphics[width=\columnwidth]{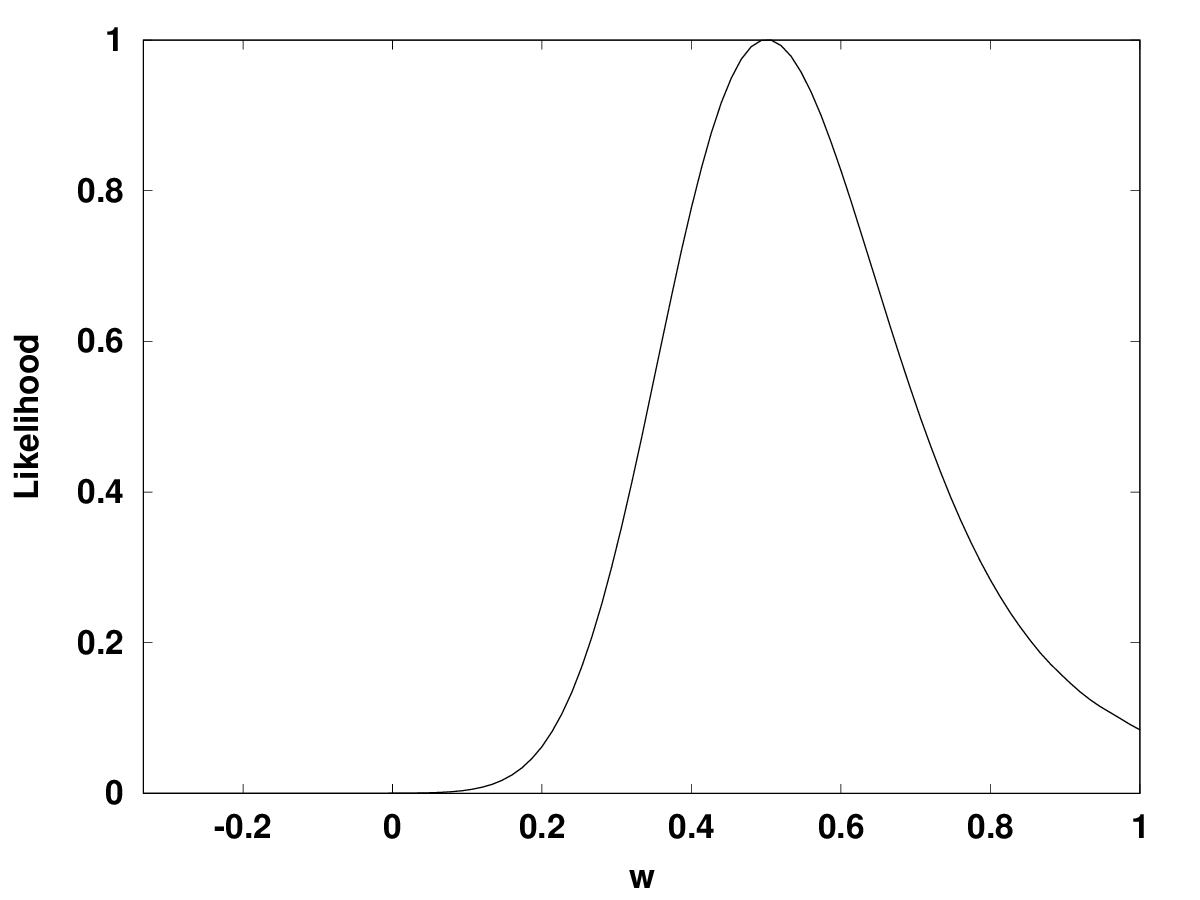}
\end{center}
\caption{\label{fig2}Constraints on the Canuto \textit{et al.} model, with $\Omega_\Lambda=0$ and $\Omega_k=0$. Left panels: Two-dimensional constraints. The black lines represent the one, two and three sigma confidence levels, and the color map depicts the reduced chi-square at each point in the parameter space, with points with $\chi^2_\nu>1.5$ shown in white. Right panels: One-dimensional posterior likelihoods for the matter density and equation of state parameters.}
\end{figure*}

We can also explore an extended parameter space allowing for a non-vanishing equation of state of the matter-like component, $w\neq0$. Specifically we assume flat uniform priors for $w\in[-1/3,1]$, and further enlarge those of the other model parameters to $\Omega_m\in[0.0, 0.5]$ and $p\in[-5, 0[$. For simplicity we again assume $\Omega_k=0$. We do note that this is mainly an academic exercise, since the equation of state of the matter component is tightly constrained, to about $|w_m|<0.003$ \cite{Thomas,Tutusaus,Ilic}. 

Figure \ref{fig2} summarizes the results of this analysis. In this extended parameter pace one can in principle get a good fit to the data (in fact the best-fit 3D set of parameters overfits the data, with a reduced chi-square of $\chi^2_\nu=0.72$), though this comes at the cost of extremely non-standard parameters. An arbitrarily small $p$ is again preferred, while the one-sigma posterior constraints on the matter density and equation of state are
\bq
\Omega_m&=&0.11\pm0.03\\
w&=&0.51\pm0.16\,,
\eq
We note that a similarly small matter density and large equation of state would be preferred in the Maeder model studied in \cite{Alternatives}. Thus, if one restricts oneself to the datasets considered in the present work, this particular model could provide a viable alternative to $\Lambda$CDM. Nevertheless, considering that the preferred values of both parameters are significantly different from the corresponding ones in canonical models, it is not at all clear that the model can still be in agreement with other datasets, e.g. the cosmic microwave background. Such an extended analysis is left for future work.

\subsection{With a cosmological constant}

In this case the models does have a $\Lambda$CDM limit, which corresponds to $p=0$. The consistency condition provided by the Friedmann equation can be written in the form
\begin{equation}
\Omega_\Lambda=\left(1+\frac{p\Omega_\lambda}{2}\right)^2-\Omega_m-\Omega_k\,,  
\end{equation}
which we will use in this subsection to eliminate $\Omega_\Lambda$.

\begin{figure*}
\begin{center}
\includegraphics[width=\columnwidth]{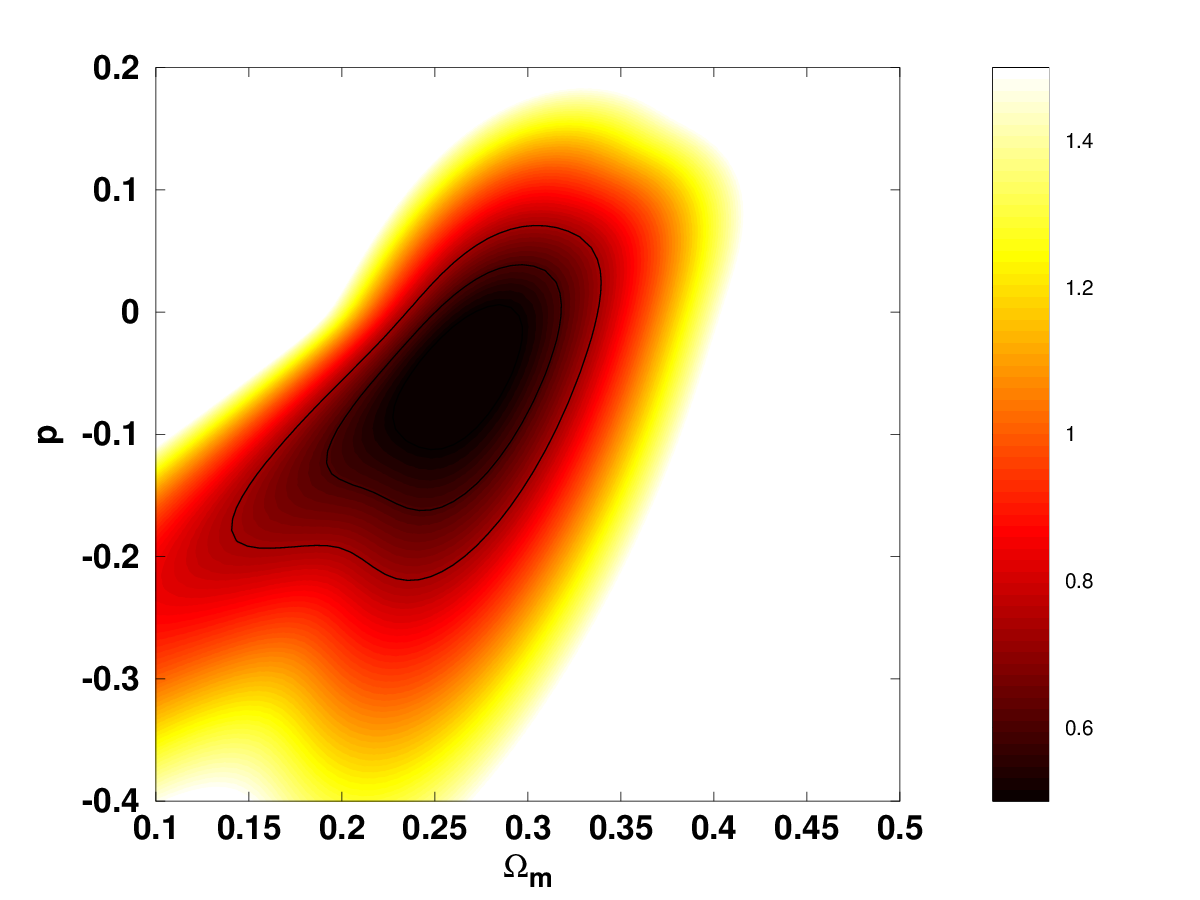}
\includegraphics[width=\columnwidth]{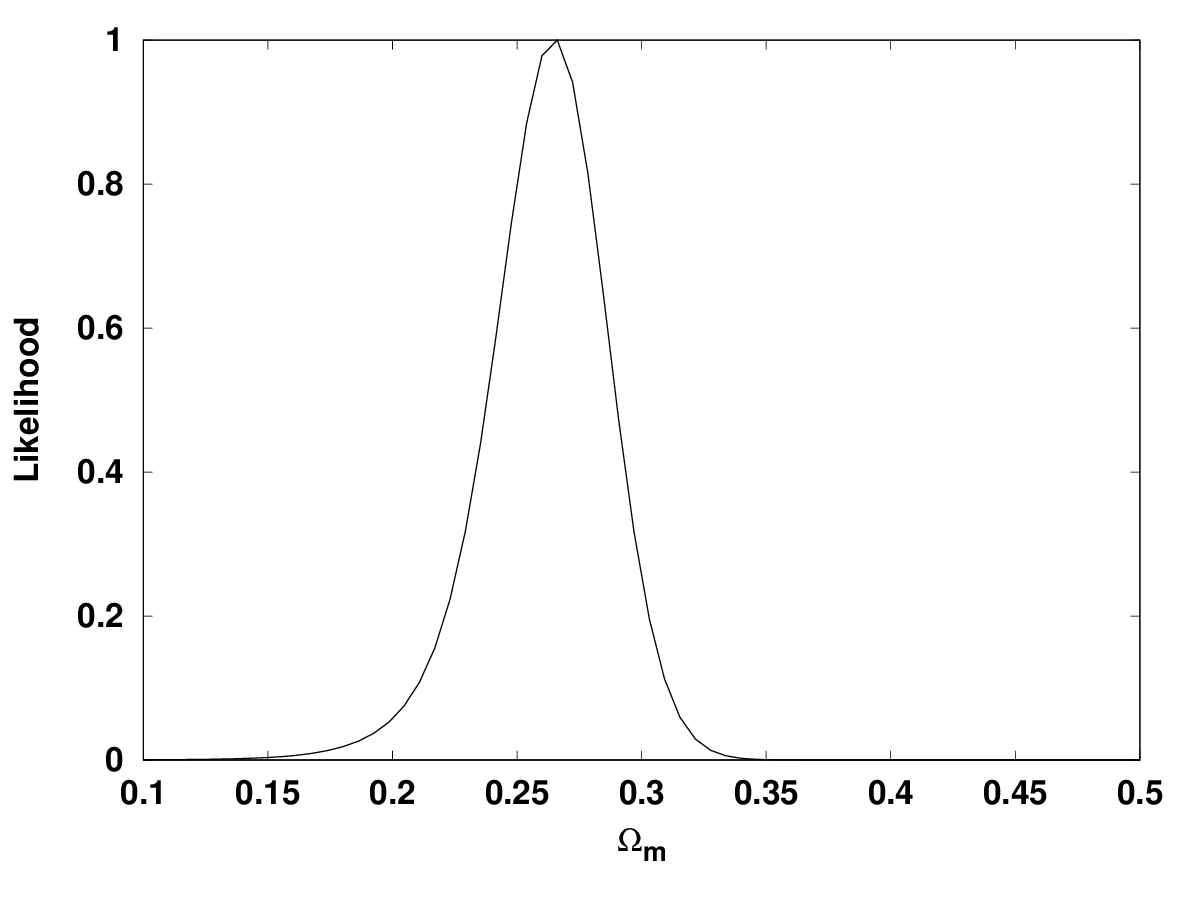}
\includegraphics[width=\columnwidth]{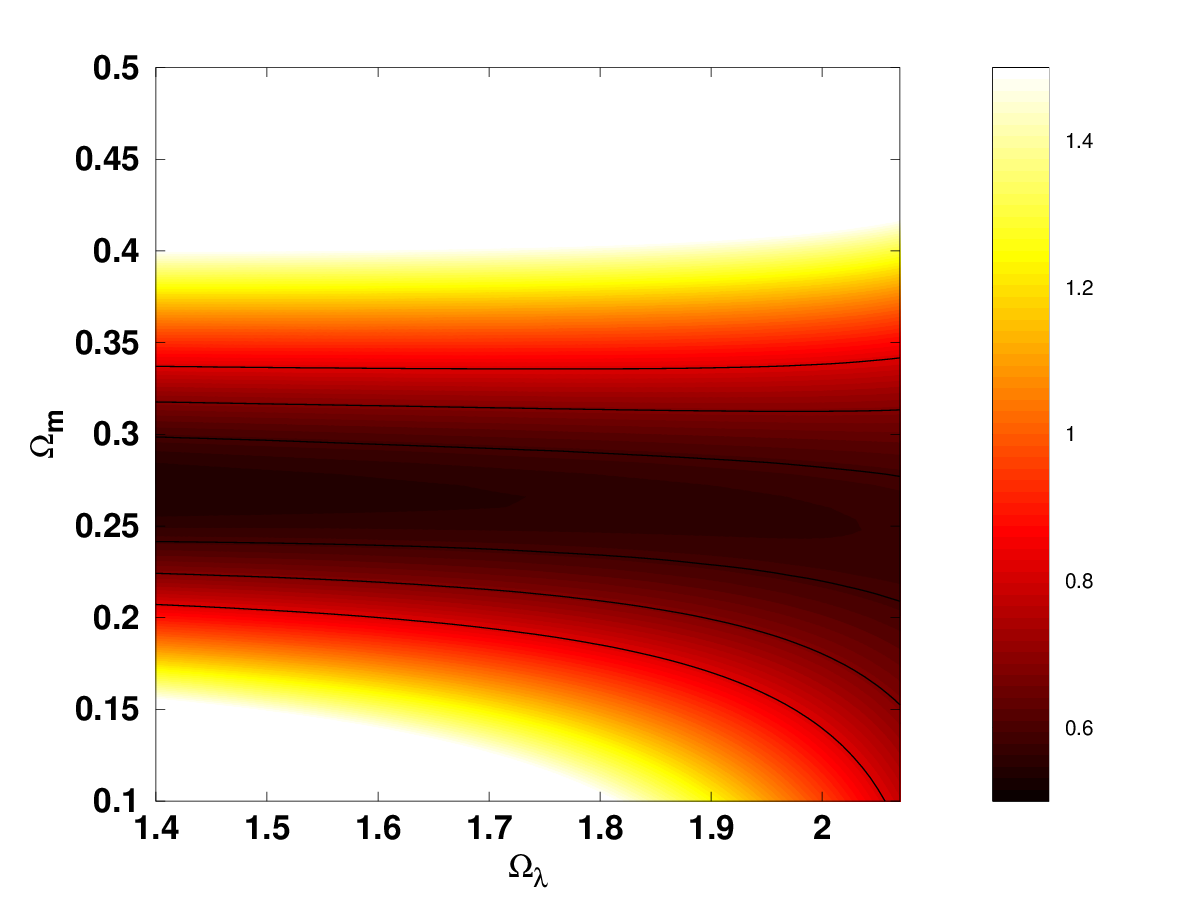}
\includegraphics[width=\columnwidth]{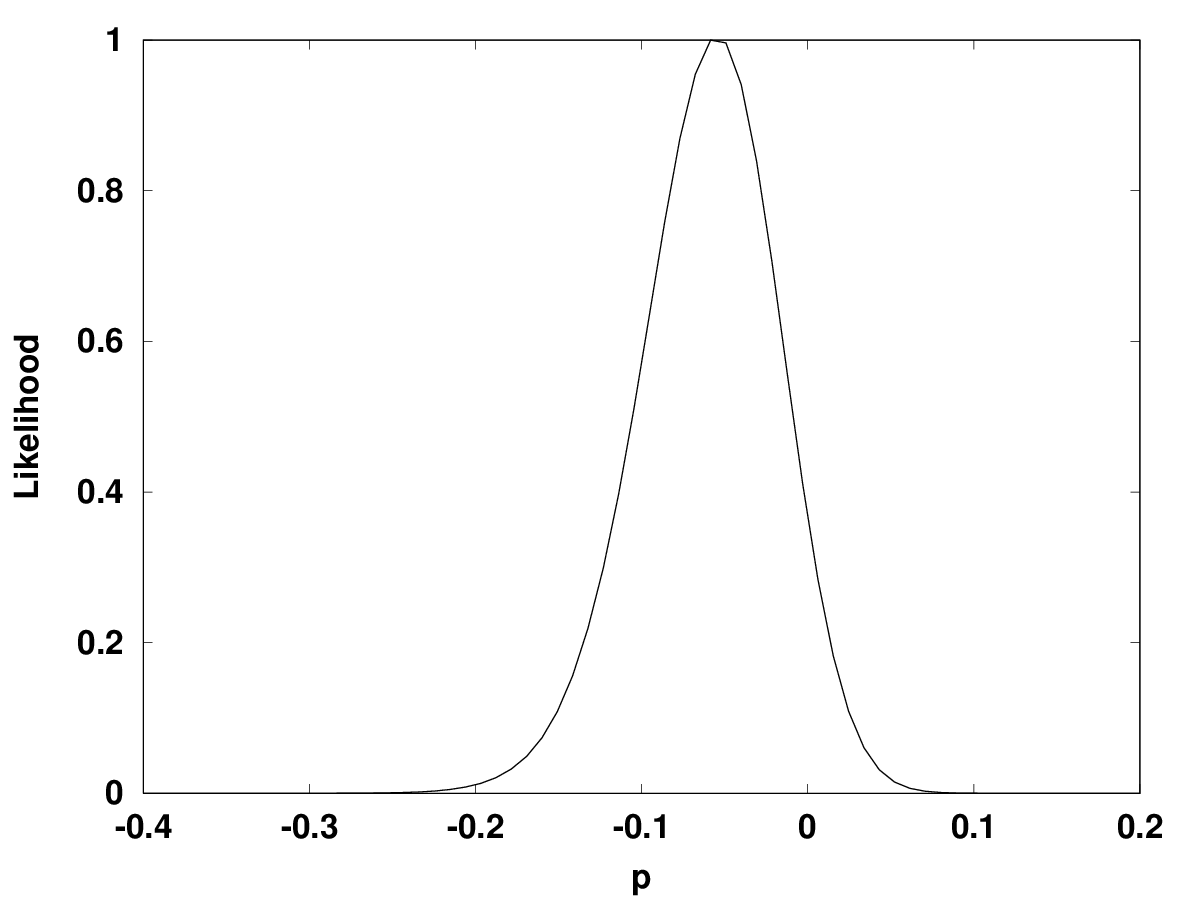}
\end{center}
\caption{\label{fig3}Constraints on the Canuto \textit{et al.} model, with $\Omega_k=0$ and $w=0$. Left panels: Two-dimensional constraints. The black lines represent the one, two and three sigma confidence levels, and the color map depicts the reduced chi-square at each point in the parameter space, with points with $\chi^2_\nu>1.5$ shown in white. Right panels: One-dimensional posterior likelihoods for the matter density and $p$ parameters.}
\end{figure*}

We start with the simplest parameter space, assuming $\Omega_k=0$ and $w=0$ together with the priors $\Omega_m\in[0.1, 0.5]$ and $p\in[-0.4, 0.2]$. As for the effective parameter $\Omega_\lambda$, which is related to the present day age of the universe, we assume the prior $\Omega_\lambda\in[1.4, 2.07]$. The upper limit corresponds to a minimum age of the Universe of 13.5 Gyr, the age of the oldest identified galaxy, GN-z11 \cite{Eleven}, while the lower limit corresponds to an age of of 20 Gyr.

Figure \ref{fig3} shows the results of this analysis. We find that the best-fit 3D set of parameters again overfits the data, with a reduced chi-square of $\chi^2_\nu=0.64$, while the one-sigma posterior constraints on the matter density and the power of the time-dependent function are
\bq
\Omega_m&=&0.27^{+0.02}_{-0.03}\\
p&=&-0.06^{+0.05}_{-0.04}\,,
\eq
while the effective parameter $\Omega_\lambda$ is not well constrained. These are of course consistent with $\Lambda$CDM.

\begin{figure*}
\begin{center}
\includegraphics[width=\textwidth]{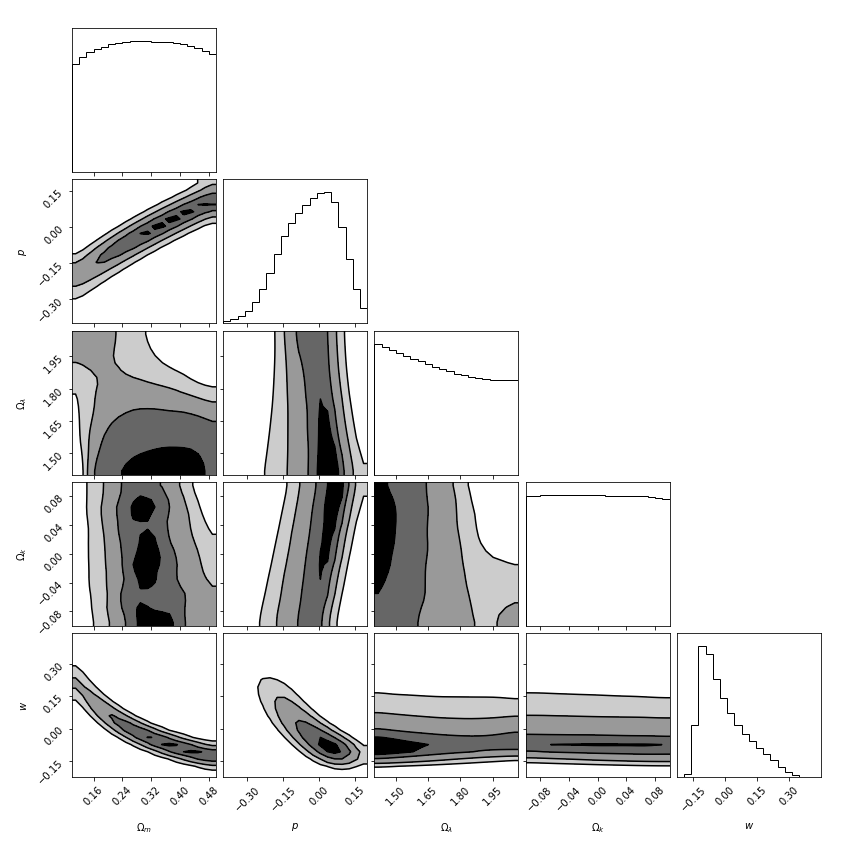}
\end{center}
\caption{\label{fig4} Corner plots for the MCMC analysis of the full five-dimensional parameter space of the Canuto \textit{et al.} model, with the priors specified in the main text. The 2D panels the depict the one, two and three sigma confidence regions.}
\end{figure*}

This parameter space already overfits the data, so from a purely statistical point of view there is no need to consider extensions allowing for curvature or a non-standard matter equation of state. Nevertheless it is worth, for completeness, to explore the full parameter space, with the goal of ascertaining whether other local maxima of the likelihood exist and identify possible degeneracies between model parameters. Towards this end we have also done an MCMC analysis of the full five-dimensional parameter space, using the same parameter priors as above, specifically $\Omega_m\in[0.1, 0.5]$, $p\in[-0.4, 0.2]$, $\Omega_\lambda\in[1.4, 2.07]$, $\Omega_k\in[-0.1, 0.1]$ and $w\in[-1/3,1]$. Figure \ref{fig4} depicts the results, and highlight the very significant degeneracies in this extended parameter space. This implies that one sigma constraints on the model parameters are significantly relaxed, to
\bq
\Omega_m&=&0.25\pm0.13\\
p&=&0.03^{+0.11}_{-0.17}\\
w&=&-0.10^{+0.23}_{-0.08}\,,
\eq
while $\Omega_\lambda$ and $\Omega_k$ are unconstrained. Nevertheless these remain, as expected, compatible with $\Lambda$CDM,


\section{Outlook}\label{sect6}

We have extended previous work \cite{Alternatives} by observationally constraining the general class of models proposed by Canuto \textit{et al.} in the context of the low redshift acceleration of the universe. These models effectively have a time-dependent cosmological constant, and depending on further assumptions may or may not have a parametric $\Lambda$CDM limit. Our analysis relied on background low redshift cosmological observations, and used constraints on the standard CPL phenomenological parameterization as a benchemark against which to compare our results.

Comparing our benchmark flat CPL model, which we have briefly discussed in Sect. \ref{sect2}, with the datasets under consideration, we find that the best-fit value of the matter density is $\Omega_m\sim0.26$ (slightly increased to $\Omega_m\sim0.27$ for the particular case of a constant equation of state parameter, $w(z)=w_0$). The scale covariant model leads to a nominally similar best-fit value of the matter density, if one assumes a flat universe and a standard matter equation of state. Allowing for curvature only increases the preferred matter density by about one standard deviation. However, if one assumes a vanishing cosmological constant these models lead to a large reduced chi-square, so they do not provide a fit to the data as good as that of the CPL model. If one insists on the assumption of a vanishing cosmological constant, the only possible way forward to get a good statistical fit would be to relax the assumption of pressureless matter. In that case one can decrease the reduced chi-square (to the extent that the data actually overfits the model), though at the cost of a low matter density ($\Omega_m\sim0.11$) and an extreme equation of state ($w\sim0.51$). These values are very different from those inferred using probes, such as the cosmic microwave background. Such a behaviour also occurs for the Maeder model, the particular model in this class that has been constrained in previous work \cite{Alternatives}.

Thus the conclusion is that these models do not provide a competitive alternative to $\Lambda$CDM. If one imposes $\Omega_\Lambda=0$, reasonable statistical fits can only be obtained at the cost of non-standard parameter values, highly discrepant from those obtained with other probes. On the other hand, if a cosmological constant is allowed, these models are parametric extensions of $\Lambda$CDM. The key model parameter describing deviations from the canonical behaviour is $p$, and as expected this is tightly constrained and compatible with the standard value.

It is worthy of note that part of the original motivation for these models stems from Dirac's Large-number Hypothesis \cite{Dirac1,Dirac2}. It follows that one further observational consequence of these models would be a time variation of Newton's gravitational constant. This has not been addressed in the present work, the main reason being that such a variation will affect the peak luminosity of Type Ia supernovas \cite{Gaztanaga,Wright}, effectively making it redshift dependent, and such effects need to be consistently included in the analysis. Similarly one may expect some effects in the Hubble parameter measurements, whether these come from cosmic chronometers or from BAO data. All these effects will need to be self-consistently included in the analysis, either through a phenomenological parametrization of the redhsift dependence of Newton's gravitational constant or through the choice of a specific model (e.g., one inspired by string theory, or a modified gravity one) where such a redshift dependence occurs. This clearly warrants a separate treatment, but in the meantime do point out that there are now stringent constraints on such variations \cite{Uzan}; these, together with early universe cosmological data (e.g., from cosmic microwave background observations) will further constrain these models.

Overall, we therefore conclude that $\Lambda$CDM is a robust paradigm. While it is clearly a phenomenological approximation to a still unknown more fundamental model, it is nevertheless a good one, and any plausible alternative model must be able to closely reproduce its behaviour in a broad range of cosmological settings.

\section*{Acknowledgements}

This work was financed by FEDER---Fundo Europeu de Desenvolvimento Regional funds through the COMPETE 2020---Operational Programme for Competitiveness and Internationalisation (POCI), and by Portuguese funds through FCT - Funda\c c\~ao para a Ci\^encia e a Tecnologia in the framework of the project POCI-01-0145-FEDER-028987 and PTDC/FIS-AST/28987/2017. 

\bibliographystyle{model1-num-names}
\bibliography{canuto}
\end{document}